%% file: main.tex
\documentclass[%
 aip,
 jcp,
 amsmath,amssymb,
 reprint,%
]{revtex4-1}

\usepackage{braket}
\usepackage{graphicx}\usepackage{dcolumn}
\usepackage{bm}

\usepackage[utf8]{inputenc}
\usepackage[T1]{fontenc}
\usepackage{mathptmx}
\usepackage{etoolbox}
\usepackage{xcolor}

\def\mbf{\mathbf}
\def\t{\text}

\def\beit{\begin{itemize}}
\def\eit{\end{itemize}}
\def\mbf{\mathbf}
\def\mbb{\mathbb}
\def\l{\left}
\def\r{\right}

\def\Ep{E_\t{P}(q)}
\def\Ec{E_\t{C}(q,m)}
\def\Em{E_\t{M}}
\def\rabi{\Omega_\t{R}}

\makeatletter
\def\@email#1#2{%
 \endgroup
 \patchcmd{\titleblock@produce}
  {\frontmatter@RRAPformat}
  {\frontmatter@RRAPformat{\produce@RRAP{*#1\href{mailto:#2}{#2}}}\frontmatter@RRAPformat}
  {}{}
}%
\makeatother
\begin{document}


\title[]{Static disorder-induced renormalization of polariton group velocity}
\author{Gustavo J. R. Aroeira}
\author{Raphael F. Ribeiro}
\email[]{raphael.ribeiro@emory.edu}
\affiliation{Department of Chemistry and Cherry Emerson Center for Scientific Computation, Emory University, Atlanta, GA, 30322}

\date{\today}

\begin{abstract}
Molecular exciton-polaritons exhibit long-range, ultrafast propagation, yet recent experiments have reported  far slower propagation than expected. In this work, we implement a nonperturbative approach to quantify how static energetic disorder renormalizes polariton group velocity in strongly coupled microcavities. The method requires no exact diagonalization or master equation propagation, and depends only on measurable parameters: the mean exciton energy and its variance, the microcavity dispersion and the Rabi splitting. Using parameters corresponding to recently probed organic microcavities, we show that exciton inhomogeneous broadening slows both lower and upper polaritons, particularly when the mean exciton energy fluctuation approaches the collective light-matter coupling strength. A detailed discussion and interpretation of these results is provided using perturbation theory in the limit of weak resonance scattering. Overall, our results support the view that exciton–phonon interactions likely dominate the recent experimental observations of polariton slowdown in disordered media. 
\end{abstract}

\maketitle
\input{a_Introduction}
\input{b_Methods}
\input{c_Results}
\input{d_Conclusions}

\vspace{0.2in}

\textbf{Data Availability.} The data that support the findings of this study are available from the corresponding author upon reasonable request. 

\vspace{0.2 in}

\textbf{Acknowledgments.} R.F.R acknowledges support from NSF
CAREER award Grant No. CHE-2340746 and start-up funds from Emory University. R.F.R. is grateful to Tal Schwartz for generously sharing  Bloch-surface wave polariton dispersion data and Milan Delor for clarifying discussions on the inhomogeneous broadening of perovskite excitons.

\vspace{0.2 in}
\textbf{References}
\bibliography{refs}%
\end{document}

%% file: a_Introduction.tex
\textbf{Introduction.} Hybrid light-matter excitations denoted polaritons \cite{hopfield1958theory, agranovich1959propagation, lidzey1998strong, agranovich2003cavity, ebbesen2016hybrid} have emerged as promising mediators of long-range, ultrafast and low-loss energy transport\cite{basko2000electronic, coles2014polariton, schachenmayer2015cavity, feist2015extraordinary, zhong2017energy, Myers2018,Schfer2019,hou2020ultralong,Botzung2020,Wei2021,Guo2022BoostingSubstrate,Allard2022,Nosrati2023,engelhardt2023polariton, SezBlzquez2018, Xiang2020IntermolecularCoupling, li2021collective} crucial for new optoelectronic technologies \cite{Nikolis2019,Wang2021, Sandik2024}. In contrast with molecular exciton transport which is typically short-ranged and diffusive \cite{zhu2019ultrafast, ginsberg2020spatially}, the hybrid light-matter character of exciton-polaritons\cite{ribeiro2018polariton, herrera2020molecular} enables ballistic energy flow in molecular materials across micrometer distances via the resonant coupling of molecular excitons to the standing waves of a confined electromagnetic field \cite{kavokin2017microcavities,Tibben2023,zhou2024nature}. 

Yet, recent ultrafast microscopy experiments revealed polariton propagation speeds far below the group velocities predicted from the polariton dispersion ~\cite{pandya2022tuning,balasubrahmaniyam2023enhanced,xu2023ultrafast, hong2025exciton}. This slowdown has been mainly attributed to dynamical disorder originating from exciton-phonon interactions which scatter polaritons and disrupt coherent wavepacket propagation~\cite{ xu2023ultrafast,sokolovskii2023multi,blackham2025microscopictheorypolaronpolaritondispersion,sokolovskii2024disentangling,chng2025quantum,Ying2025}. In support of this interpretation, temperature-dependent measurements in perovskite microcavities\cite{xu2023ultrafast} show that transport velocities approach the expected group velocity as the system approaches $T \to 0~\mathrm{K}$, where thermal phonon populations vanish and inelastic phonon-assisted scattering is suppressed. Further corroboration of the important role played by phonon-assisted scattering in polariton transport was provided by semiclassical simulations including intra and intermolecular exciton-phonon interactions ~\cite{sokolovskii2023multi,xu2023ultrafast, chng2025quantum, krupp2024quantum, blackham2025microscopictheorypolaronpolaritondispersion}. 

\par Yet even in the absence of phonons, polariton transport could remain unexpectedly slow. Static disorder arising from sample inhomogeneities\cite{ziman1979models} introduces elastic scattering that limits coherent propagation. Our recent theoretical works examining coherent transient exciton propagation in polaritonic wires showed that, in fact, static disorder can significantly slow down exciton-polariton transport \cite{aroeira2023theoretical, aroeira2024coherent} (see also \cite{Allard2022, engelhardt2023polariton}). However, static disorder is most impactful in one-dimensional systems \cite{mott1961theory, ishii1973localization, abrahams1979scaling, ping2006mesoscopic}, and the typical magnitude of static disorder-induced changes in polariton group velocity in higher dimensions remain unknown. 

\par In this work, we directly address this open question. We employ a nonperturbative method \cite{litinskaya2006loss, litinskaya2008propagation} to quantify the effects of static exciton disorder on polariton group velocities in organic microcavity exciton-polaritons. Our results establish the typical magnitude of polariton group velocity renormalization induced by resonant scattering. We find static disorder leads to potentially significant slower transport mainly in lower polariton modes at large in-plane wave-vectors, where the exciton content is high. Strong renormalization emerges only when the mean exciton energy fluctuation approaches half the Rabi splitting. Perturbative estimates support these trends and provide analytical insight into the numerically obtained behavior. 

%% file: b_Methods.tex
\par \textbf{Methods.} We model a large, isotropic
and homogeneous molecular ensemble under strong coupling with a microcavity with perfectly reflective mirrors. In this regime, polariton energies $\Ep$ (with $P = \t{LP}$ or $\t{UP}$), can be approximated as solutions to an effective medium dispersion relation originally derived in the context of molecular polaritons by Litinskaya and Reineker \cite{litinskaya2006loss} and that can be expressed as
\begin{align}   & \Ep - \Ec =\frac{\rabi^2}{4} \int_{-\infty}^{\infty}dE' \frac{\rho(E')}{\Ep - E'}
    \label{eq:res_scatter1},
\end{align}
where $\Ec$ is the energy of the photon mode with in-plane wave-vector magnitude $q = \sqrt{q_x^2+q_y^2}$,  and longitudinal wave-vector $k_z(m) =m\pi/L_\t{C}$, with $m \in \mathbb{N}_{>0}$ for transverse-electric and $m \in \mathbb{N}_0$ for transverse-magnetic modes,  $\rabi$ is the Rabi splitting, and the probability distribution $\rho(E)$ models the exciton static disorder (inhomogeneous broadening). We assume $\rho(E)$ is a normal distribution with mean $E_M$ and variance $\sigma^2/2$, \begin{align} \rho(E)=\frac{1}{\sigma\sqrt\pi}\exp\l[-(E-E_M)^2/\sigma^2\r]. \end{align} With this choice, Eq. \ref{eq:res_scatter1} can be written as
\begin{align}   & \Ep - \Ec =\frac{\rabi^2\sqrt{\pi}}{4i\sigma}e^{-\frac{[\Ep -\Em]^2}{\sigma^2}}\text{erfc}\left[\frac{\Ep- \Em}{i\sigma}\right] \label{eq:res_scatter2},
\end{align}
where erfc is the complementary error function \cite{abramowitz1965handbook}. Due to the fast growth of erfc$(-ix)$ for large $x\in \mbb{R}$, it is numerically advantageous to work instead with the scaled complementary error function $\text{erfcx(z)} = \text{exp}\left(z^2\right)\text{erfc}(z)$, where $z = -i[E(q)-E_M]/\sigma$ as this substitution preserves floating-point accuracy at large $|z|$.

\par The solutions to Eq. \ref{eq:res_scatter2} are complex-valued polariton energies $E_P(q)$ for any nonzero disorder strength $\sigma > 0$, with the imaginary part characterizing the finite polariton lifetime due to elastic resonant scattering  \cite{hopfield1969resonant, lagendijk1996resonant, litinskaia2002elastic}. Static disorder also shifts the real part of $E_{P}(q)$ and leads to the renormalized group velocity 
\begin{align} v_g(q) =\frac{1}{\hbar}\frac{\partial}{\partial q} \t{Re}[\Ep].\label{eq:vgq}\end{align} 
As discussed in detail in  Refs.~\cite{litinskaya2006loss, suyabatmaz2023vibrational}, this approach is suitable so long as the polariton spectral function $A(\mbf{q},E) = \braket{\mbf{q}|\delta(E-H)|\mbf{q}}$ (where $\mbf{q}$ corresponds to the microcavity mode with in-plane wave vector $\mbf{q}$) is narrowly centered around $\t{Re}~E_P(q)$. In other words, Eq. \ref{eq:vgq} provides a non-perturbative estimate of the renormalized polariton group velocity if the disorder-induced broadening $\delta  q$ satisfies $\delta q/q < 1$, where $\delta q$ can be estimated from the imaginary part of the renormalized polariton energy \cite{agranovich2003cavity,litinskaya2006loss, suyabatmaz2023vibrational}
\begin{align}
    \delta q = \frac{\t{Im}[\Ep]}{\hbar v_g(q)}\;. \label{dq}
\end{align}

%% file: c_Results.tex
\begin{figure}[h]
    \includegraphics[width=\columnwidth]{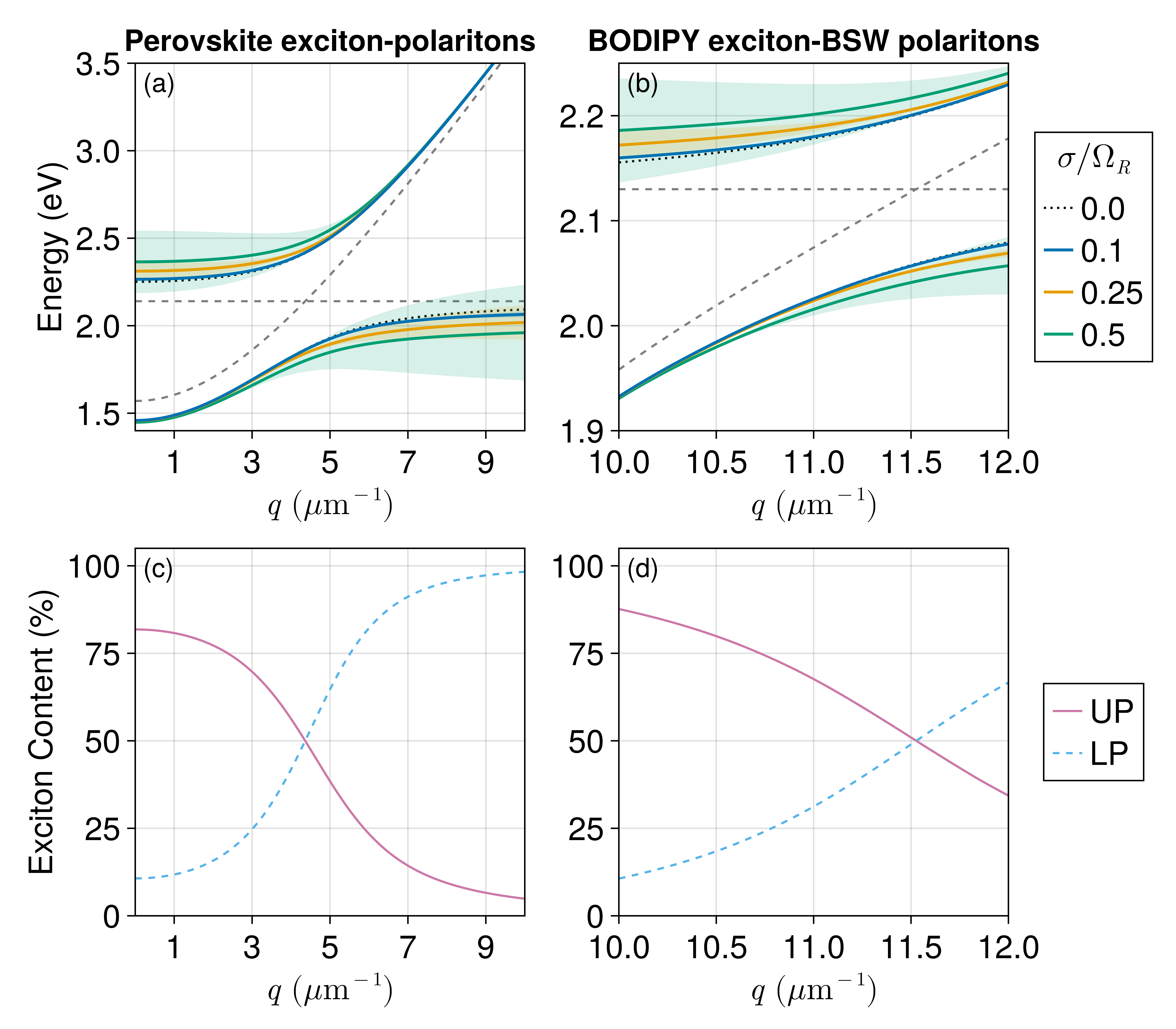}
    \caption{Polariton dispersion relations obtained from Eq. \ref{eq:res_scatter2} using parameters characterizing light and matter systems experimentally studied by (a) Xu et al. \cite{xu2023ultrafast}  and (b) Balasubrahmanyam et al.  \cite{balasubrahmaniyam2023enhanced}. Band plots represent energy broadening $\pm\t{Im}[E_P(q)]$ arising from resonant scattering. Horizontal vertical and curved dashed lines represent the bare exciton ($\Em$) and photon ($\Ec$) energies.  Panels (c) and (d) show the estimated exciton content for each polariton mode with in-plane wave vector $q$ in systems (a) and (b), respectively.} \label{fig:dispersion}
\end{figure}

\textbf{Results and Discussion.} Figures \ref{fig:dispersion}(a) and (b) show renormalized polariton dispersion relations obtained by solving Eq. \ref{eq:res_scatter2} using light-matter parameters from recent studies of perovskite-microcavity polaritons\cite{xu2023ultrafast} and BODIPY exciton-Bloch surface wave polaritons \cite{balasubrahmaniyam2023enhanced}, respectively (see Table \ref{tab:parameters}). To provide a complete description of the static disorder influence on the polariton group velocity, we simulated the renormalized dispersion of the aforementioned systems under vanishing, weak and strong disorder strengths ($\sigma/\Omega_R = 0.0,\; 0.1,\;0.25,\;0.5$). For the perovskite system examined in Ref. \cite{xu2023ultrafast}, a two-dimensional spectroscopy study by Kandada \textit{et al.}  suggests $\sigma = 5 -10~\text{meV}$\cite{srimath2022homogeneous} which corresponds to $\sigma/\Omega_R < 0.2$ \cite{xu2023ultrafast}. \\

\par Figure \ref{fig:dispersion} shows the effect of static disorder on the dispersion of the examined systems is similar: LP frequencies are reduced, whereas UP frequencies are blueshifted. In the LP branch, the effect is vanishingly small at $q$ close to 0  and becomes more relevant as $q$ becomes larger. Conversely, in the UP branch, static disorder effects are maximal at $q \rightarrow 0$ and become negligible at large $q$. These trends can be understood based on the exciton content of each polariton mode as presented in Figs. \ref{fig:dispersion}(c) and (d). At $q \rightarrow 0$, the LP modes are mostly photonic and therefore are negligibly affected by matter static disorder, whereas the UP has much greater exciton content, and the corresponding mode frequencies are much more sensitive to static disorder. The same reasoning holds at large $q$, where the increased exciton content of LP modes yields a greater energy renormalization. 

\begin{table}[]
    \centering
    \renewcommand{\arraystretch}{1.5}
    \begin{tabular}{|c|c|c|}
        \hline
         Parameter & (a)\cite{xu2023ultrafast} & (b)\cite{balasubrahmaniyam2023enhanced} \\
         \hline
         Bare exciton energy $(E_M)$ & 0.214 eV & 0.213 eV  \\
         Lowest photon energy $(E_C)$ & 0.157 eV & N/A \\
         Rabi splitting  $(\rabi)$ & 0.550 eV & 0.142 eV \\
         Microcavity length $(L_C)$ & 0.667 $\mu$m & N/A \\
         \hline
    \end{tabular}
    \caption{Light and matter parameters used in this work corresponding to (a) Perovskite exciton-microcavity polaritons\cite{xu2023ultrafast} and (b) BODIPY exciton-BSW polaritons \cite{balasubrahmaniyam2023enhanced}. All parameters including the dispersion of the BSW system were obtained from the cited works.}
    \label{tab:parameters}
\end{table}

\par The renormalization of the polariton dispersion by static disorder presented in Fig. \ref{fig:dispersion} is only significant when $\sigma$ approaches $\rabi/2$, and even then, only the frequencies of modes with high exciton content ($> 50 \%$) are substantially shifted. Observing these variations would require polariton homogeneous linewidths to be smaller than the change in energy induced by disorder-induced elastic scattering. To verify if that is the case, the dispersion curves in Figs. \ref{fig:dispersion}(a) and (b) include shaded regions representing the energy uncertainty $\pm \t{Im}[\Ep]$ imparted by resonance scattering. This uncertainty is generated by the exciton inhomogeneous broadening and neglects additional broadening mechanisms such as microcavity photon leakage and dynamical disorder which further increase the polariton energy width. Nevertheless, there is already significant overlap between the zero-disorder and the broadened renormalized polariton dispersions obtained with the considered relative disorder strengths ($\sigma/\Omega_R$), therefore suggesting that the renormalized dispersion is unlikely to be observable in linear optical response measurements under typical experimental conditions.

\par The static disorder-induced lowering of LP and raising of UP energies can be understood from a weak disorder perturbative expansion of Eq. \ref{eq:res_scatter2} in powers of $\sigma/\rabi$. For modes with $\delta q/q \ll 1$ , in the limit where $\sigma/\rabi \rightarrow 0$,  $|\Ep - \Em|/\sigma \rightarrow \infty$, and we can employ the asymptotic expansion \cite{arfken2011mathematical}
\begin{align}
    \t{erfc}(z) \sim \frac{e^{-z^2}}{z\sqrt{\pi}}\l[\frac{1}{z} - \frac{1}{2z^3} + O\l(z^{-5}\r) \r] \;, ~~ z \rightarrow \infty \label{eq:assympt}
\end{align}
Inserting the first two terms of Eq. \ref{eq:assympt} into Eq. \ref{eq:res_scatter2} leads to the following weak disorder renormalized dispersion relation valid to $O(\sigma^2/\Omega_R^2)$
\begin{align}
   [\Ep - \Ec][\Ep - \Em]-\frac{\rabi^2}{4} = 
    \frac{\rabi^2 \sigma^2}{8[E^{(0)}_\t{P}(q) - \Em]^2} \;,
\end{align}
where $E^{(0)}_\t{P}(q)$ is the polariton energy obtained from Eq. \ref{eq:res_scatter1} in the limit where $\sigma \rightarrow 0$ and we used $\Ep =E^{(0)}_\t{P}(q) +O\l(\sigma^2\r)$. The leading change in polariton energies induced by weak disorder can be written as
\begin{align}
      [\Ep - \Ec][\Ep - \Em] -\frac{\tilde{\Omega}_\t{R}^2}{4} \approx 0, \label{eq:mod_disp}
\end{align}
where 
\begin{align}
    \tilde{\Omega}_\t{R} \approx \rabi\sqrt{1+ \frac{\sigma^2}{2\left[E^{(0)}_\t{P}(q) - \Em \right]^2} } \;. \label{eq:modifiedrabi}
\end{align}
It follows that $\tilde{\Omega}_\t{R} > \rabi$; hence, static disorder increases the splitting between the LP and UP branches, i.e., $E_\t{LP}(q) < E_\t{LP}^{(0)}(q)$ and $E_\t{UP}(q) > E_\t{UP}^{(0)}(q)$. This simple analysis also explains the asymmetric renormalization effect on polariton dispersion seen in Fig. \ref{fig:dispersion}. The change in energy is greatest for a given $q$ at the branch $\t{P}$ with smaller $|\Ep - \Em|$. In redshifited microcavities (where $E_C(0) < E_M$ as in both systems considered in Fig. \ref{fig:dispersion}), this perturbative treatment holds for UP modes with large $q$ and LP modes with small $q$, where $|\Ep - \Em| \gg \sigma$. As $\Ep$ approaches the bare exciton energy (UP modes with small $q$ and LP modes with large $q$), the weak disorder assumption breaks down. However, these modes have $\delta q/q \geq 1$ and therefore, Eq. \ref{eq:res_scatter2} is no longer applicable. 

\par From the previous discussion on the renormalization of the polariton dispersion relation, we can deduce static disorder reduces both lower and upper polariton group velocities as we explain next. Let the renormalized polariton energies be written as $\Ep  = E^{(0)}_P(q) + \hbar\delta_P(q)$, where $\delta_P(q)$ is the static disorder induced energy shift. The corresponding renormalized group velocities are given by $v_\t{P}(q) = \hbar^{-1}\partial \Ep/\partial q = v_\t{P}^{(0)}(q) +\delta'_\t{LP}(q)$, where $\delta'_\t{LP}(q) =  \partial \delta_\t{P}(q)/\partial q$. As $q$ grows, $\sigma/|E_\t{LP}(q) - \Em|$ increases resulting in an enhanced polariton gap and greater in magnitude LP redshift (see Eq. \ref{eq:modifiedrabi}). This leads to the conclusion that $\delta'_\t{LP}(q) <0$ and therefore static disorder induces slower LP propagation, i.e., $v_\t{LP}(q) < v_\t{LP}^{(0)}(q)$. 
For the UP branch, as $q$ increases $\tilde{\Omega}_\t{R} \rightarrow \rabi$ (see Eq. \ref{eq:modifiedrabi}). Hence, as shown by Fig. \ref{fig:dispersion}, the effect of static disorder on the UP dispersion subsides at high $q$, i.e., $\delta_\t{UP}(q) \rightarrow 0$. Given that $\delta_\t{UP} \geq 0$ (from Eqs. \ref{eq:mod_disp} and \ref{eq:modifiedrabi}) and that $\delta_{\t{UP}}$ becomes smaller as $q$ increases, we conclude that $\delta'_\t{UP}(q) < 0$ and $v_\t{UP}(q) < v_\t{UP}^{(0)}(q)$. In summary, the provided perturbative arguments are highly suggestive that static disorder generically reduces polariton group velocities.

\par
\begin{figure}[h]
    \includegraphics[width=\columnwidth]{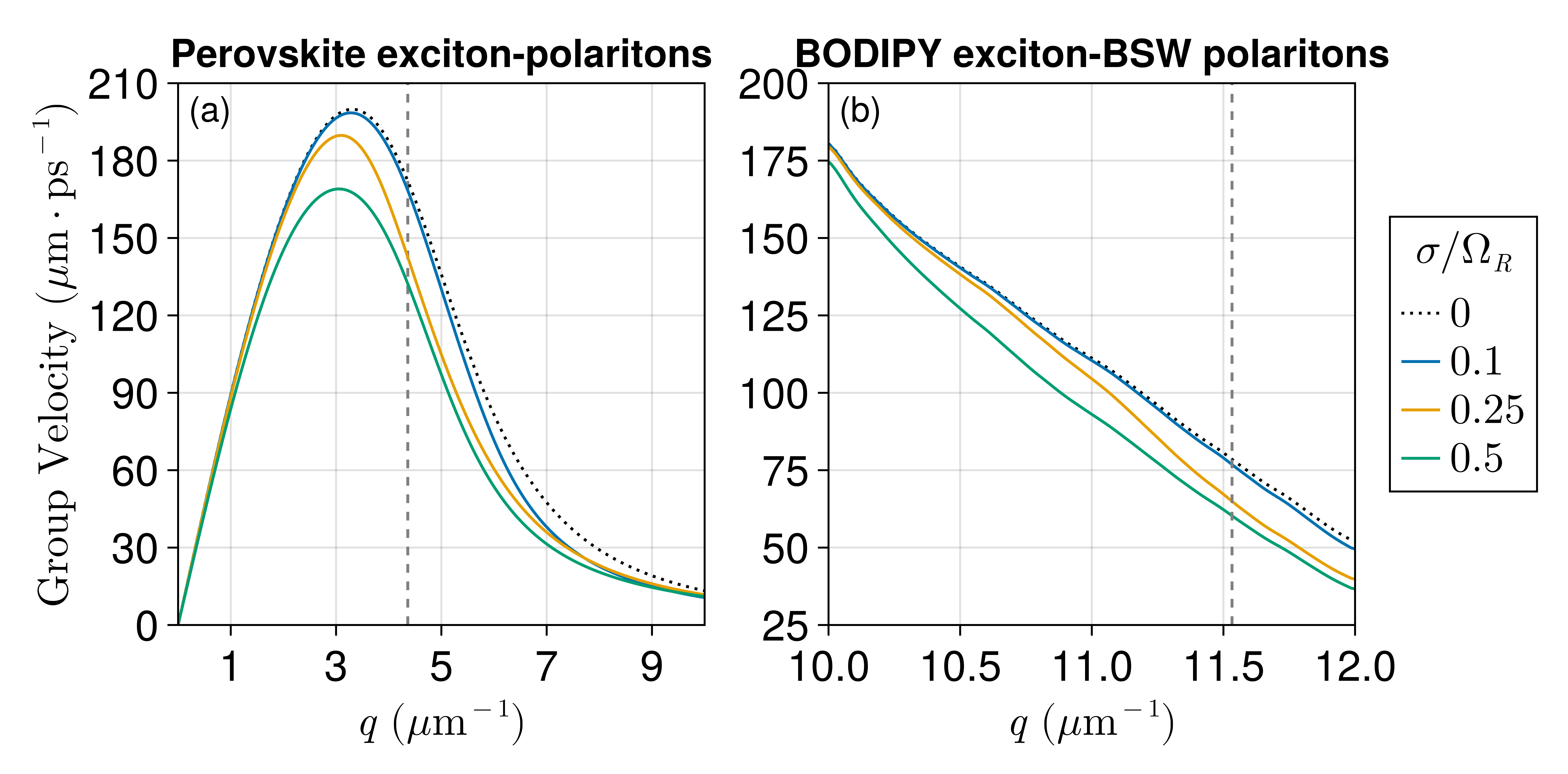}
    \caption{Renormalized LP group velocities obtained from numerical differentiation of the dispersion curves in Fig. \ref{fig:dispersion} as determined by Eq. \ref{eq:res_scatter2} with light-matter parameters from (a) the perovskite exciton-polariton system examined in Ref. \cite{xu2023ultrafast} and (b) BODIPY exciton-Bloch surface wave polaritons from Ref.  \cite{balasubrahmaniyam2023enhanced}. The vertical dotted line marks the resonance point where $E_C(q) = E_M$.}\label{fig:gv}
\end{figure}

 In Fig. \ref{fig:gv}, we present numerical verification of the reduction in polariton group velocity induced by static disorder for the same systems analyzed in Fig. \ref{fig:dispersion}. The results, obtained from Eqs. \ref{eq:vgq} and \ref{eq:res_scatter2}, support the argument that static disorder consistently lowers the group velocity across all polariton branches with the largest reduction occurring at the subset of wave vectors where polaritons are predominantly molecular in character. Strong disorder effects appear only when $\sigma$ is comparable to $\Omega_R$, and they mainly affect excitations with exciton content above 50$\%$. 

\par Because substantial intermolecular interactions protect polaritons from localization \cite{bradbury2024stochastic, hong2025exciton}, the actual group velocity renormalization due to static disorder in exciton-polaritons arising in organic semiconductors with strong dipolar coupling is likely even weaker than what is shown here. Hence, our results support the perspective that at finite temperatures (or low temperatures in systems with strong vibronic coupling \cite{krupp2024quantum}) dynamical disorder arising from phonon scattering likely dominates over static effects. This conclusion is consistent with temperature-dependent group velocity measurements by Xu et al \cite{xu2023ultrafast}.

\par
\begin{figure}
    \centering
    \includegraphics[width=0.9\linewidth]{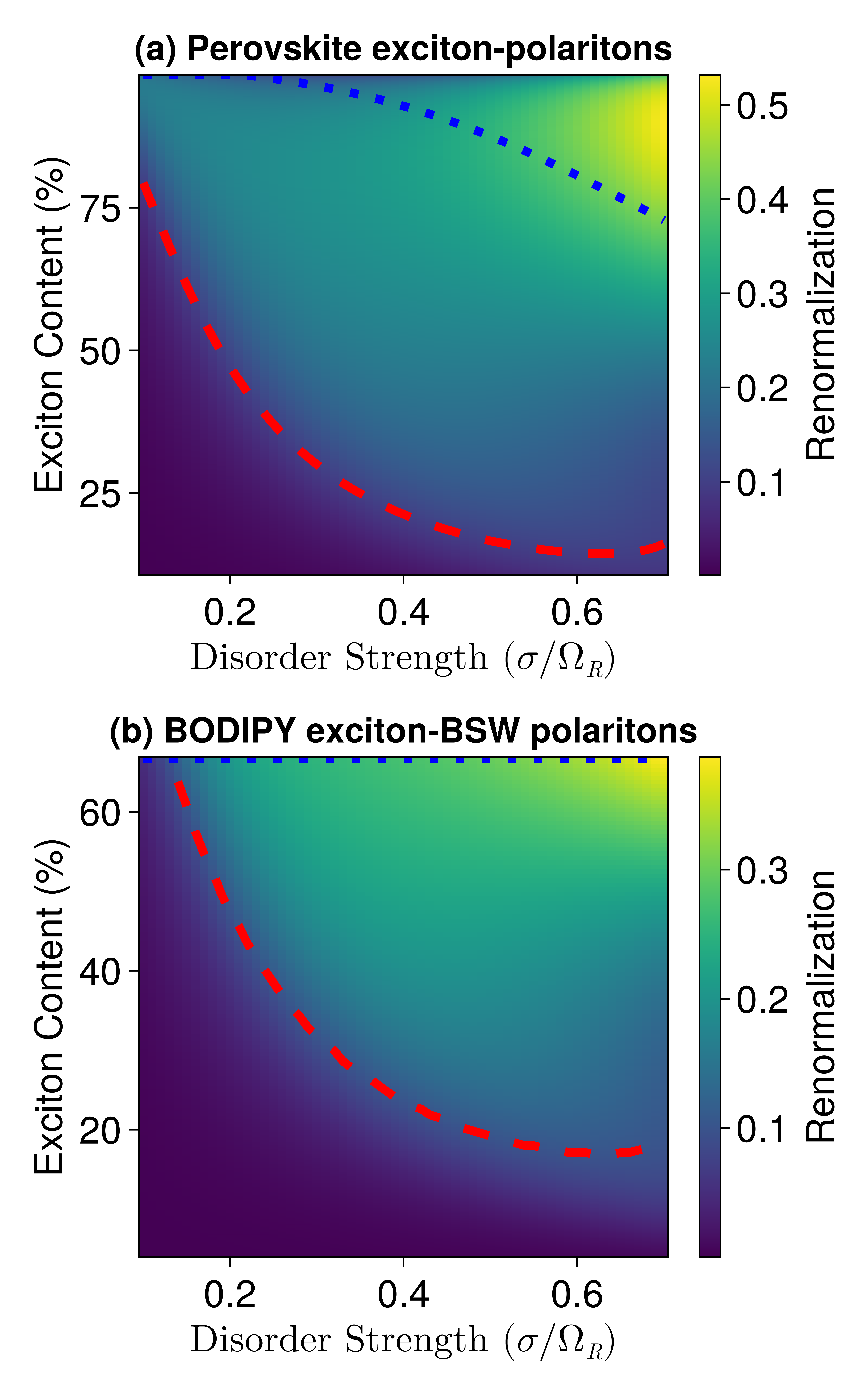}
    \caption{LP Group velocity renormalization as function of exciton content and static disorder strength relative to $\rabi$. Bare light and matter parameters are obtained from (a) Xu et al. \cite{xu2023ultrafast} and  (b) Balasubrahmanyam et al. \cite{balasubrahmaniyam2023enhanced}. The red dashed line marks the boundary over which static disorder-induced renormalization becomes greater than 10\%. The blue dotted line delimits the region where polaritons are expected to be well characterized by their in-plane wave vector magnitude $q$, i.e., $\delta q / q < 1$.}
    \label{fig:heatmap}
\end{figure}

 Figure~\ref{fig:heatmap} compiles our key findings, showing how static disorder renormalizes the LP group velocity as a function of the relative disorder strength, $\sigma/\rabi$, and the exciton fraction $P_M(q)$. We quantify this renormalization with the dimensionless quantity $1 - v^{(0)}_{g}(q)/v_{g}(q)$, where $v^{(0)}_{g}(q)$ is the LP group velocity at in-plane wave vectors with magnitude $q$ in the absence of static disorder and $v_{g}(q)$ its disorder-renormalized counterpart. The red dashed contour marks the 10$\%$ renormalization threshold, while the blue dotted contour signals the onset of substantial wave vector broadening, $\delta q/q > 1$, beyond which $q$ is unlikely to be an approximately conserved quantity. Panels~\ref{fig:heatmap}(a) and (b) confirm the intuitive trend that less disorder is needed to slow  polaritons when its exciton character is larger. Quantitatively, the crossover boundary between negligible and significant velocity renormalization (as defined by the 10$\%$ renormalization criterion) follows approximately the relation $P_M(q) \propto (\sigma/\rabi)^{-1}$ for $\delta q/q < 1$.

\par Overall, even at strong static disorder, $\sigma/\Omega_R > 0.5$, the computed velocity reduction remains modest compared with the values reported in Refs.~\cite{xu2023ultrafast,balasubrahmaniyam2023enhanced}. This finding reinforces the view that phonon-assisted  scattering, rather than static disorder, dominates the experimentally observed molecular exciton-polariton group velocity renormalization under typical conditions.

\par Taken together with the preceding analysis, Fig.~\ref{fig:heatmap} supports the view that static disorder can slow polaritons appreciably only when they are strongly excitonic and no longer possess a well-defined in-plane wave vector. Under typical experimental conditions at moderate temperatures, static disorder therefore plays, at most, a supporting role compared with dynamical disorder arising from exciton–phonon scattering.



%% file: d_Conclusions.tex
\par \textbf{Conclusions.} We implemented a nonperturbative framework \cite{litinskaya2006loss} to estimate how resonance scattering induced by static energetic disorder renormalizes molecular polariton group velocities. This approach requires only the probability density function for the exciton transition energy, the collective light-matter interaction strength as provided by the Rabi frequency, and the bare microcavity dispersion, so it can be readily applied to a wide range of material platforms. 

\par Our numerical simulations employing parameters corresponding to the polaritonic systems experimentally probed in Refs.~\cite{xu2023ultrafast,balasubrahmaniyam2023enhanced} reveal that static disorder can explain at most a minor fraction of the observed group velocity renormalization. Thus, we are led to conclude most of the reported polaritonic slowdown stems from dynamical (phonon-assisted) disorder \cite{sokolovskii2024disentangling}. This mechanism is likely dominant in most molecular exciton-polariton systems, especially in the presence of significant vibronic coupling and moderate temperatures, where inelastic scattering induced by exciton-phonon interactions dominates over elastic scattering due to static fluctuations. Nevertheless, at low temperatures, the thermal phonon population dwindles and elastic scattering induced by a frozen disordered environment takes over. Under these conditions, materials exhibiting strong static energetic disorder and weak vibronic coupling will show polariton group velocity renormalization mirroring the behavior reported here.